\documentclass[aps,prl,reprint,groupedaddress]{revtex4-1}

\usepackage{stackengine,graphicx,xcolor}
\usepackage{amsmath}
\usepackage{etaremune}
\usepackage{multirow}

\begin{document}

\title{Indistinguishable Single-Photon Sources \\with Dissipative Emitter Coupled to Cascaded Cavities}
\author{Hyeongrak Choi$^1$}
\email[]{choihr@mit.edu}
\author{Di Zhu$^1$}
\author{Yoseob Yoon$^2$}
\author{Dirk Englund$^1$}
\email[]{englund@mit.edu}
\affiliation{$^1$Research Laboratory of Electronics, Massachusetts Institute of Technology, Cambridge, Massachusetts 02139, USA}
\affiliation{$^2$Department of Chemistry, Massachusetts Institute of Technology, Cambridge, Massachusetts 02139, USA}
\date{\today}

\begin{abstract}
Recently, Grange et al. [Phys. Rev. Lett. 114, 193601 (2015)] showed the possibility of single photon generation with high indistinguishability from a quantum emitter, despite strong pure dephasing, by `funneling' emission into a photonic cavity. Here, we show that cascaded two-cavity system can further improve the photon characteristics and greatly reduce the $Q$-factor requirement to levels achievable with present-day technology. Our approach leverages recent advances in nanocavities with ultrasmall mode volume and does not require ultrafast excitation of the emitter. These results were obtained by numerical and closed-form analytical models with strong emitter dephasing, representing room-temperature quantum emitters. 
\end{abstract}

\maketitle

Sources of indistinguishable single photons play an essential role in quantum information science~\cite{Igor.Solid}, including linear-optics quantum computing~\cite{Kok.Linear,Terry.GHZ,Mihir.Percolation}, precision measurements~\cite{Seth.Sensing}, quantum simulation~\cite{Aspuru.Random}, boson sampling~\cite{Aaronson.BS,Spring.Boson}, and all-optical quantum repeaters~\cite{Azuma.QR,Mihir.Rate}. Single photon sources based on atom-like quantum emitters have seen remarkable progress~\cite{Igor.Solid, Senellar.SPS}, including in particular color centers in diamond, many of which have been shown to possess long spin coherence times. However, a remaining challenge is to improve their emission properties to achieve near-unity indistinguishability and high collection efficiency. 

Here we show that the emitter decay can be tailored by coupling to a cascaded two-cavity system, which provides enough control to minimize pure dephasing and spectral diffusion. Our analysis shows that the cascaded-cavity scheme improves on the photon emission efficiency ($\eta$) and indistinguishability ($I$) compared to previously considered single-cavity approaches. For the especially difficult problem of room-temperature operation with silicon vacancy centers in diamond, the cascaded-cavity system enables same efficiency, but much higher indistinguishability ($\sim0.95$) than the single cavity case ($\sim0.80$), with $\times20$ lower cavity quality factor ($Q$-factor). When the cavities are tuned for maximum $\eta I$ product, more than two orders of magnitude improvement compared to the bare emitter, and a $\sim17\%$ improvement over the best single-cavity system are possible.

\begin{figure}
\includegraphics[width = \columnwidth, trim=5 5 5 5,clip]{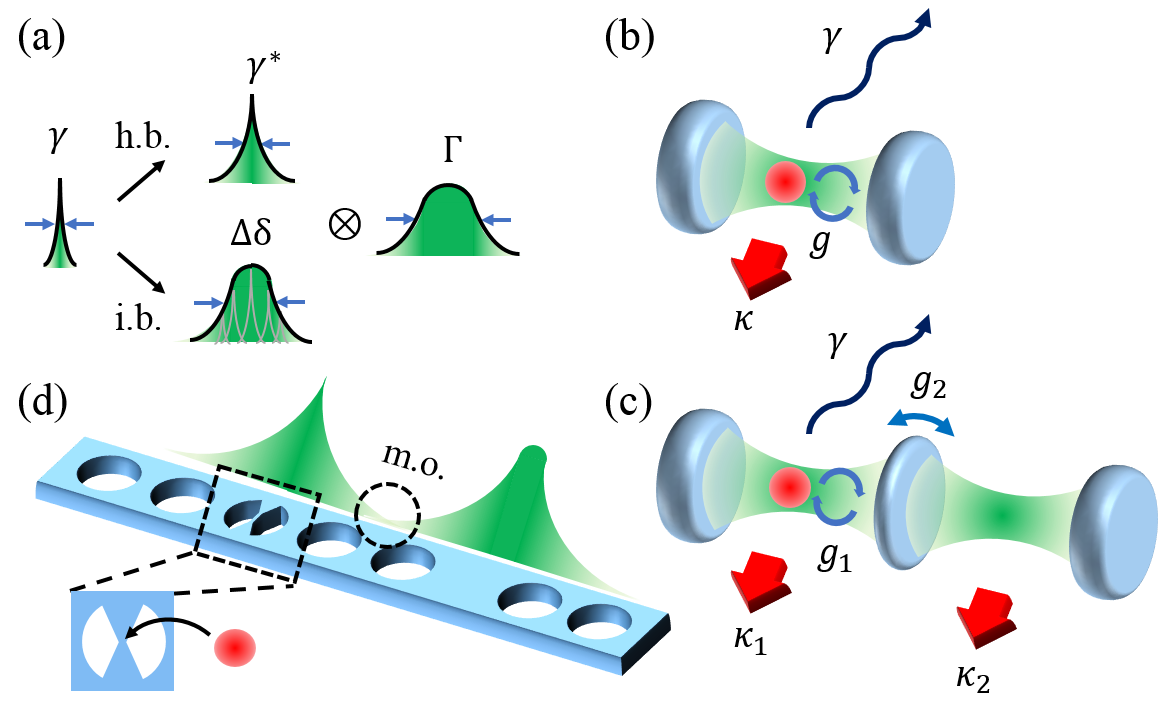}
\caption{(a) The emission spectrum with width $\Gamma$ can be considered as a convolution of different broadenings: natural broadening (Lorentzian linewidth $\gamma$), pure dephasing (Lorentzian linewidth $\gamma^*$), and spectral diffusion (Gaussian linewidth $\Delta \delta$). (b) Cavity QED system, where $g$ is the coupling rate, $\kappa$ is the cavity decay rate, and $\gamma$ is the spontaneous emission rate of the emitter to non-cavity modes. (c) Cascaded cavity system as a room temperature single photon source. An emitter is coupled to the first cavity ($C_1$) with coupling strength $g_1$. $C_1$ is coupled to the second cavity ($C_2$) with coupling strength $g_2$. $\kappa_1$ and $\kappa_2$ are the cavity radiation losses to free space. (d) Photonic crystal realization of the proposed cascaded-cavity-emitter system. The first cavity produces a high emitter-cavity coupling  ($g_1$) due to field concentration in a concentric dielectric tip, see~\cite{Weiss.ACS,Chuck.PRL}. A mode overlap (m.o.) of cavities corresponds to a cavity-cavity coupling rate of $g_2$.}
\end{figure}

As shown in Fig. 1.(a), the linewidth of an emitter is given by $\Gamma = \gamma + \gamma^* + \Delta \delta \gg \gamma$ at room temperature, where $\gamma$ is the radiative decay rate, $\gamma^{*}$ the pure dephasing rate, and $\Delta \delta$ the FWHM width of the spectral diffusion. Pure dephasing can be modeled as Markovian phase flip process that occurs much faster than the excited-state lifetime. Whereas $\Delta \delta$ captures the spectral diffusion between photoemission events (for example due to changing stray electric fields near the emitter) that changes much slower than the excited-state lifetime; thus, spectral diffusion can be treated by a statistical average over the ensemble. The indistinguishability, which is approximately given by $I\sim\gamma/\Gamma$~\cite{Wong.Ind}, is vanishingly small at room temperature ($\sim10^{-4}$ for silicon vacancy centers in nanodiamond~\cite{Bechor.Temp}). 

Nanophotonic structures have been investigated to improve $I$ by modifying the local density of electromagnetic states (LDOS). This approach can be analyzed in its simplest form in the cavity quantum electrodynamics (cavity-QED) picture of Fig. 1(b), where $g\propto 1/\sqrt{\text{V}_\text{eff}}$ the emitter-cavity coupling rate, $\text{V}_\text{eff}$ the cavity mode volume, $\kappa\propto 1/Q$ the cavity decay rate, and $Q$ the quality factor. For simplicity, we first ignore spectral diffusion, i.e., $\Gamma=\gamma+\gamma^*$. In the incoherent regime, where the $\Gamma+\kappa\gg 2g$, the system dynamics reduces to a set of rate equations, in which the emitter and the cavity field pump each other at rate $R = 4g^2/(\Gamma+\kappa)$~\cite{SI}.

There are two main approaches to increase $I$. One strategy is to maximize the LDOS with plasmonic cavity, so that $R=4g^2/(\Gamma + \kappa) > \gamma^*$~\cite{Fre.Plasmon, Simon.Plasmon}. Peyskens et al. showed that for a $20$ nm silver nanosphere ($Q\sim 15$) coupled to a waveguide, the indistinguishability of single photons emitted from SiV can be increased to $I\sim0.27$ while reaching a single photon out-coupling efficiency of $\eta\sim0.053$. On the other hand, Wein et al. could theoretically achieve $I\sim0.37$ and $\eta\sim0.77$ with the plasmonic Febry-Perot hybrid cavity ($Q\sim986$) recently proposed in~\cite{Gurlek.Hybrid}. However, this approach also faces several important obstacles: (1) the assumption of instantaneous pumping on the femtoseconds scale, which is demanding due to ionization (resonant) and slow phonon relaxation (non-resonant)~\cite{Turro.Text}; (2) and Ohmic and quenching losses in the metal. 

A second approach investigated by Grange et al.~\cite{Thomas.Funnel} relies on coupling the emitter to a dielectric cavity with ultrahigh $Q$, which avoids the problems of high losses in metals. When the cavity decay rate $\kappa$ is much smaller than $\gamma$ and $R$, near-unity indistinguishability becomes possible. Notably, this system outperforms the spectral filtering of a emitted photon, due to a `funneling' of emission into the narrow-band cavity spectrum. However, reaching an indistinguishability of 0.9 (0.5) for an emitter with $\gamma\sim 2\pi\times100$ MHz radiative linewidth at $\omega\sim 2\pi\times 400$ THz requires a cavity with very high $Q\sim4 \times10^{7 (6)}$; this $Q$ far exceeds the highest quality factor nanocavity coupled to a quantum emitter, which has $Q\sim 55,000$~\cite{Arakawa.QD}. The underlying problem is that high indistinguishability is not possible with the limited $Q$ and $V_\text{eff}$ that are currently available.  

The cascaded two-cavity system considered in this paper, illustrated in Fig. 1(c), greatly reduces the $Q$ factor requirements while obtaining higher overall single photon source performance. The emitter is assumed to be dipole-coupled with the first cavity ($C_1$). This cavity can have a relatively low $Q$  factor $<10^5$, as long as it has a small $V_\text{eff}$ to efficiently collect the emitter fluorescence. However, the indistinguishability $I$ of the emission from cavity $C_1$ would be low. A high $I$ can then be achieved by coupling to a second cavity ($C_2$), which provides additional degrees of freedom to optimize the single photon emission from the two-cavity-emitter system. 

To investigate the dynamics quantitatively, we assume a strong pure dephasing, $\gamma^*=10^4$, normalized to $\gamma=1$. In the regime where the total dephasing ($\Gamma$) exceeds the emitter-$C_1$ coupling rate $g_1$, the population transfer rate between the emitter and $C_1$ becomes~\cite{SI},
\begin{align}
R_1 = \frac{4g_1^2}{\Gamma+\kappa_1}\cdot\frac{1}{1+(\frac{2\delta}{\Gamma+\kappa_1})^2},
\end{align}
where $\delta$ is the detuning, assumed to be $0$ for now. A large transfer rate $R_1>\gamma$ (implying $g_1\gg 1$) is required for efficient emission into the cavity. To this end, we make use of a new cavity design using a dielectric concentrator in a photonic crystal (PhC) nanocavity (Fig. 1(d))~\cite{Weiss.ACS,Chuck.PRL}. This nanocavity enables arbitrarily small mode volume~\cite{Chuck.PRL}; indeed, recently $V_\text{eff} = 10^{-3} (\lambda/n_\text{Si})^3$ was experimentally demonstrated in a silicon PhC~\cite{Weiss.Exp} with a quality factor of $\sim10^{5}$. Here, we consider $g_1=500$ and $\kappa_1=50$, corresponding to $V_\text{eff} = 0.007(\lambda/n_\text{diamond})^3$ and $Q\sim50$k for the case of silicon vacancy centers in the diamond~\cite{Jelena.Cavity}. In addition, the emitter is assumed to be located at the narrow bridge section of this cavity design, where the phonon environment is similar with the nanodiamond~\cite{Bechor.Temp}.

\begin{figure}
\includegraphics[width = \columnwidth, trim=10 5 10 10,clip]{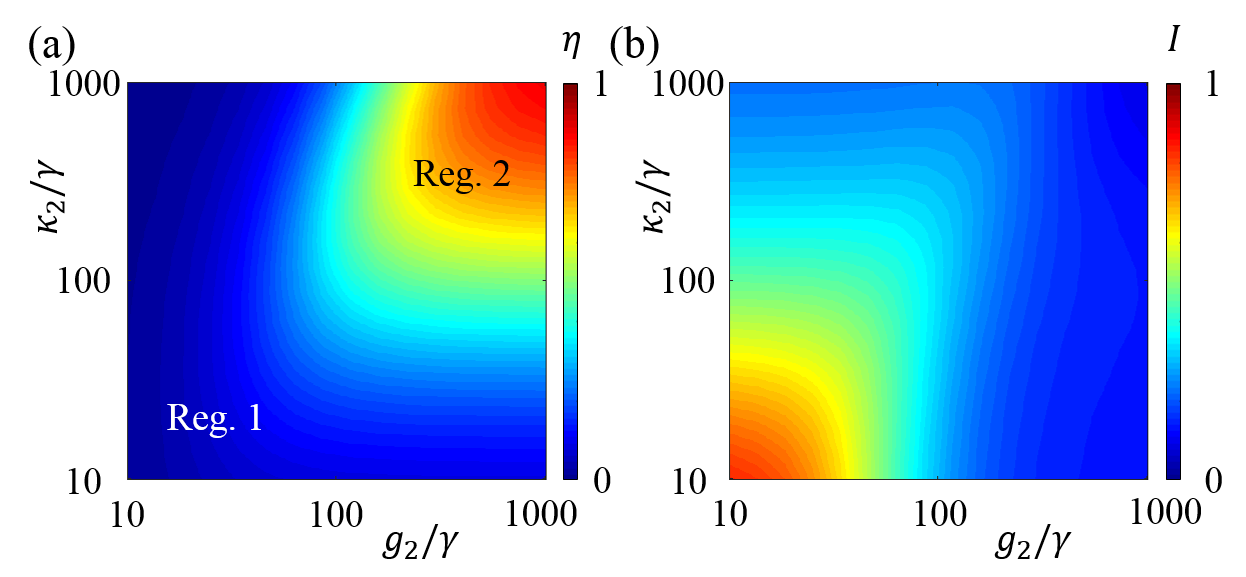}
\caption{Performance of the cascaded cavity system with ($g_1=500$, $\kappa_1=100$) as a function of $g_2$ and $\kappa_2$. Efficiency (a) and indistinguishability (b).}
\end{figure}

The second cavity ($C_2$) is coupled to $C_1$ at rate $g_2$. This coupling can be adjusted through the spacing between the two PhC cavities, as shown in Fig. 1(d). We derived the population transfer rate between cavity $C_1$ and $C_2$ from the optical Bloch equations, by applying adiabatic elimination of the coherence between the cavities in the limit of $R_1+\kappa_1+\kappa_2 \gg 2g_2$ (see derivation in the Supplemental~\cite{SI}). The population transfer rate between $C_1$ and $C_2$ then becomes
\begin{align}
R_2 = \frac{4g_2^2}{R_1+\kappa_1+\kappa_2},
\end{align}
where $\kappa_2$ is the decay rate of $C_2$. Note that if $R_1\rightarrow 0$, $R_2$ reduces to the Purcell-enhanced emission rate. The $R_1$ term in the denominator effectively acts as additional decoherence. The reduction of the population transfer rate due to this additional decoherence was studied previously in the classical and continuous-wave (CW) limits, including for non-resonant excitation of a quantum dot~\cite{Alexia.PRB} and for light transmission in an optomechanical system~\cite{Tang.Cascaded}. However, this decoherence directly impacts the indistinguishability. Thus, we investigate the \textit{temporal dynamics} of the system using a master equation.

We applied the master equation approach to calculate $I$ and $\eta$ as a function of $g_2$ and $\kappa_2$ for our cascaded-cavity-emitter system. The results in Fig.~2 show two regimes of interest. In `Reg. 1' of $R_2, \kappa_2<\kappa_1$, we find high $I$ and small $\eta$. `Reg. 2' of $R_2, \kappa_2>\kappa_1$ leads to moderate $I$ and large $\eta$. Analogous regimes were analyzed for single cavity-QED system~\cite{Thomas.Funnel}. The photon collection efficiency into $C_2$ follows from the Bloch equations~\cite{SI} for both regimes, giving:
\begin{align}
\eta = \frac{\kappa_2R_2}{\kappa_1(\kappa_2+R_2)+\kappa_2R_2}. \label{eq_eta}
\end{align}

We first focus on Reg. 1. When $R_1\gg\kappa_1,\gamma$, the emitter and $C_1$ serves as a `composite emitter' with decoherence rate $R_1$. This effective emitter decaying with rate $\sim\kappa_1/2$ is coupled to $C_2$ with $\sim\kappa_2<\kappa_1$ at a rate $R_2/2<\kappa_1$ (coupling is asymmetric, and see supplemental~\cite{SI} for more details). We were able to derive an analytical form for the indistinguishability with the non-equilibrium Green's function for the emitter-cavity system:
\begin{align}
I = \frac{\kappa_1/2+(\kappa_2||R_2)/2}{\kappa_1/2+\kappa_2+3R_2/2}, \label{eq_ind}
\end{align}
where $\kappa_2||R_2 = \kappa_2R_2/(\kappa_2+R_2)$. The same result can be derived from the quantum regression theorem~\cite{Simon.Plasmon}. Note that this equation has the similar form with the one-cavity case~\cite{Thomas.Funnel} under the substitution $(\kappa_1/2,\kappa_2,R_2/2) \rightarrow (\gamma,\kappa,R)$ --- i.e., we can consider the $C_1$-emitter system as a `composite emitter' inside cavity $C_2$. Figure 3(a) plots $\eta$ and $I$ as a function of $\kappa_2$. Equations (3) and (4) show excellent agreement with the numerical simulations with the master equation. Notice that when $R_2+\kappa_2 \sim R_1$, $I$ exceeds the prediction from Eq. (4). Deviation of result from the prediction is more evident when $R_1$ is smaller (Fig. 3(b)). This deviation occurs because the contribution of the coherence between cavities ($\rho_{ab}(t)$) to the two-time correlation function of the cavity field, $\left<b^\dagger(t+\tau)b(t)\right>$, is not negligible~\cite{SI}. 

Next, we investigate Reg. 2 ($R_2,\kappa_2>\kappa_1$), for which large $\eta$ and moderate $I$ are possible. In the $C_1$-emitter system with $R_1 > \kappa_1$, the excitation incoherently hops back and forth between the emitter and the cavity. Therefore, $C_1$ decoheres quickly at rate $R_1$, resulting in a low $I$. On the other hand, $\kappa_1>R_1$ also results in low $I$ because the timing jitter of initial incoherent feeding exceeds the cavity lifetime. The solution is to choose $R_1>\kappa_1$ and keep the population of $C_1$ low, preventing the photon to be re-absorbed by the emitter. $C_2$ provides this additional functionality with two knobs: $R_2$ and $\kappa_2$. When $R_2, \kappa_2>\kappa_1$, excitation quickly pass through the $C_1$, resulting in low population of $C_1$ ($P_{c1}$). At the same time, the decoherence of $C_1$ at rate $R_1\cdot P_{c1}$ can be suppressed by a factor of $(R_1+\kappa_1)/(R_1+\kappa_1+R_2||\kappa_2)$. 

\begin{figure}
\includegraphics[width = \columnwidth,trim=10 5 10 10,clip]{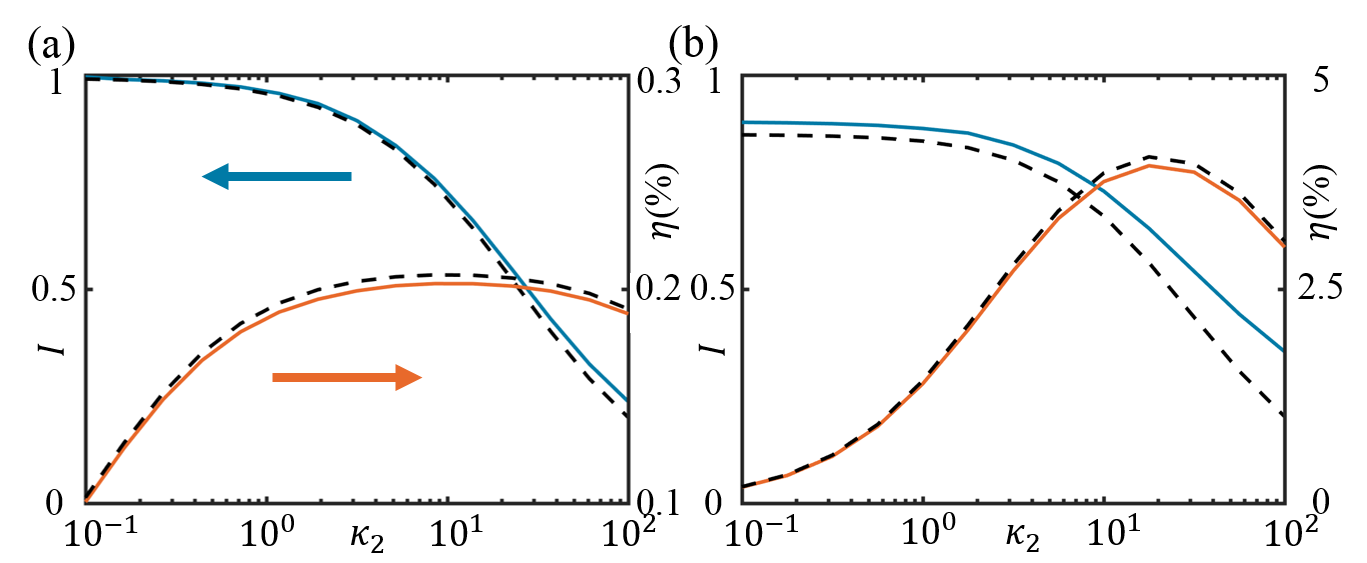}
\caption{High indistinguishability regime (Reg. 1). The emitter and $C_1$ can be treated as an effective emitter coupled to $C_2$ with population transfer rate $R_2/2$. (a) $I$ (blue) and $\eta$ (orange) as a function of $\kappa_2$ for $(g_1,\kappa_1 ,g_2) = (1500,50,5)$. (Solid) numerical result from master equation. (Dashed) analytical result from Eq.~\ref{eq_ind} ($I$) and Eq.~\ref{eq_eta} ($\eta$). (b) $I$ and $\eta$ vs. $\kappa_2$ for $(g_1,\kappa_1 ,g_2) = (500,50,10)$. The deviation between the numerical and analytical results are due to the finite effective dephasing $R_1$.}
\end{figure}

\begin{figure}
\includegraphics[width = \columnwidth,trim=12 5 5 10,clip]{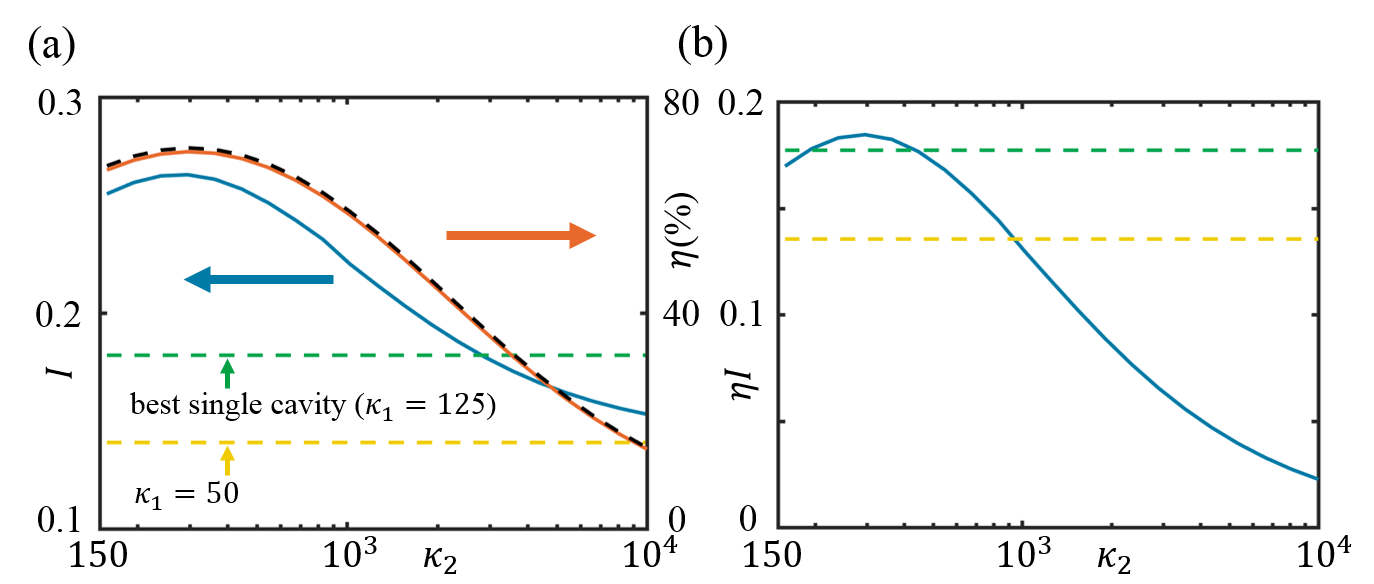}
\caption{Large $\eta I$-product regime with $(g_1, \kappa_1, g_2) = (500, 50, 150)$. (a) $I$ (blue) and $\eta$ (orange) as a function of $\kappa_2$. Black dashed line is the analytical result from Eq. (\ref{eq_eta}). For a single cavity system, $I$ is plotted with yellow dashed ($g = 500$, $\kappa = 50$) and green dashed ($g = 500$, $\kappa = 125$) lines. The latter gives the maximum $\eta I$ of single cavity systems. (b) $\eta I$-product as a function of $\kappa_{2}$ (blue). Cascaded cavity architecture shows higher $\eta I$ than that of single cavity system with $\kappa = 50$ (yellow dashed) and $\kappa = 125$ (green dashed, maximum $\eta I$).
}
\end{figure}

Figure 4(a) plots $I$ and $\eta$ of the photon emitted by $C_2$ as a function of $\kappa_2$, assuming ($g_1,\kappa_1,g_2$) =($500,50,150$). In the limit of large $\kappa_2$ and subsequently small $R_2$, the dynamics of the emitter and $C_1$ are the same without $C_2$; $C_2$ merely samples the photons from $C_1$. Thus, photons of $C_2$ have the same $I$ as that of $C_1$ without $C_2$ (single-cavity system). Simulation of single-cavity system shows that photons emitted by $C_1$ has $I\sim0.14$ (shown as a yellow dashed line). Decreasing $\kappa_2$ (increasing $R_2$) suppresses the population of $C_1$, resulting in larger $I$. Since $C_2$ suffers the same incoherent hopping/jitter effect as $C_1$ discussed above, $I$ is maximized to $I=0.27$ at $\kappa_2=300$. Notably, $I$ for a $\kappa_2=300$ exceeds the maximum achievable $I$ for a single-cavity system with $g=g_1$ (green dashed line), corresponding to the same $V_\text{eff}$. Though the improvement of $\eta I$ is less significant because of a reduced efficiency (Fig. 4(b)), $\eta I$ is still higher than the single-cavity system allows. 

Table 1 compares the $\eta$ and $I$ values achievable for the single- and cascaded-cavity architectures, assuming a silicon vacancy center in diamond at room temperature with $(\gamma,\gamma^*,\omega)\sim(160\text{ MHz}, 400\text{ GHz}, 400\text{ THz})$~\cite{Bechor.Temp}, as a quantum emitter. To achieve $I$ of $\sim 0.95$ in Reg. 1, the single-cavity approach requires a very high $Q$ factor of 50M, which is technologically challenging, especially considering integration with the emitter. The cascaded-cavity system requires only $Q_1=7$k for the first cavity and $Q_2=500$k for the second cavity to achieve the same $I$. Reaching $I\sim0.8$ requires only $Q_1=3.6$k and $Q_2=500$k for the cascaded-cavity system, whereas $Q=10$M is needed for the single-cavity system. Note that in both cases, the cascaded cavity system also achieves much higher $\eta$ values than the single-cavity case. In Reg. 2, the highest $\eta I$ is found under the constraint of $Q_\text{max}=500$k, and the cascaded-cavity system achieves $\sim17~(18)\%$ improvement of $\eta I~(I)$ over the single-cavity system.

\begin{table}[b]
\caption{Comparison of two systems} 
\begin{ruledtabular}
\begin{tabular}{l c c}
&Cascaded-cavity &Single-cavity \\
&$I$, $\eta$ ($\%$), $Q_1$, $Q_2$ &$I$, $\eta$ ($\%$), $Q$ \\
\hline

\multirow{2}{*}{Reg. 1}&
0.950, 0.76, 7k, 500k~\footnote{$g_1=500$, $\kappa_1=360$, $g_2=30$, $\kappa_2=5$}&0.950, 0.25, 50M~\footnote{$g=1.33$, $\kappa=0.05$}\\&
0.805, 3.09, 3.6k, 50k~\footnote{$g_1=500$, $\kappa_1=700$, $g_2=87.5$, $\kappa_2=50$}&0.800, 0.27, 10M~\footnote{$g=1.33$, $\kappa=0.25$}\\

\hline

Reg. 2& 
0.315, 98.6, 500k, 2.1k~\footnote{$g_1=500$, $\kappa_1=5$, $g_2=530$, $\kappa_2=1200$}&
0.267, 99.5, 3.75k~\footnote{$g=500$, $\kappa=667$}
\end{tabular}
\end{ruledtabular}
\end{table}

Finally, we statistically incorporate inhomogeneous broadening (spectral diffusion). Figures 5(a) and 5(b) plot $I$ and $\eta$ for Reg. 1 and Reg. 2 for different $\Delta\delta$ with fixed $\gamma^*=10^4$. We used a Gaussian probability distribution to model $\delta$. The key observation here is that $\Delta\delta$ does not strongly diminish $\eta$ and $I$ when $\Delta\delta\ll\gamma^*$ (note that the vertical axis is highly magnified to show detail). This insensitivity to spectral diffusion follows from Eq. (1), which shows that the transfer rate ($R_1$) is only reduced by $\sim2(\delta/\gamma^*)^2$, and subsequently from Eq. (2), that the $R_2$ changed by small amount. Suppression of the effect of small spectral diffusion only happens when the emitter is highly dissipative, i.e., if $\gamma+\gamma^*>\kappa$, and is a unique feature of the `cavity funneling' process. In other words, pure dephasing serves as a resource to stabilize the single photon source, maintaining potentially high $I$ and $\eta$ despite spectral diffusion. This result is in contrast to the bare-emitter case, where the spectral diffusion directly affects $I$~\cite{Wong.Ind}. 

\begin{figure}
\includegraphics[width = \columnwidth, trim=5 5 5 5,clip]{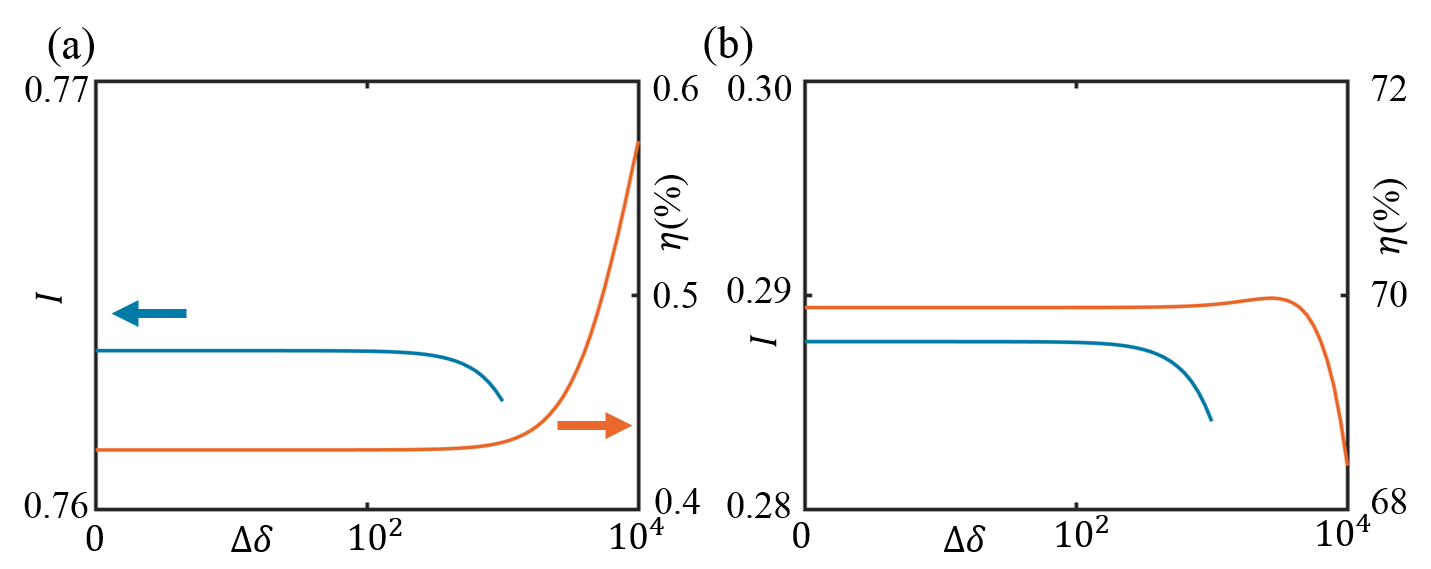}
\caption{$I$ (blue) and $\eta$ (red) in the presence of spectral diffusion. (a) Reg. 1 ($g_1,\kappa_1,g_2,\kappa_2$)=($500,50,3,10$). (b) Reg. 2 ($g_1,\kappa_1,g_2,\kappa_2$)=($500,50,150,300$). In both regimes, spectral diffusion marginally affects the $\eta$ and $I$. Note that the $y$-axes in the figures are highly magnified to see the small change across the spectral diffusion.} 
\end{figure}

We emphasize the difference of our approach with `photonic molecule' studied in~\cite{Photonic.Molecule1,Photonic.Molecule2}. In our cascaded-cavity system, the emitter only couples with mode of $C_1$, and the population is transferred between two cavity modes by weak coupling. In contrast, for the emitter coupled with photonic molecule, the emitter is coupled to two super-modes that results from strong coupling between two cavities, i.e., a splitting greater than the individual cavities' decay rates. Cascaded-cavity system is also different with the hybrid-cavity system~\cite{Gurlek.Hybrid}, where the cavity mode is modified by another cavity. More specifically, in hybrid cavity, the latter cavity field acts as a electromagnetic environment (radiation bath) of the former cavity rather than as a independent cavity mode.

In conclusion, we analyzed funneling through a cascaded-cavity scheme. We showed the two-cavity scheme dramatically reduces the requirements on the cavity $Q$-factor to the point that today's devices appear feasible. We analyzed the two-cavity-emitter system through numerical simulations, which we were able to match with the closed-form analytical solutions. By incorporating pure dephasing and spectral diffusion, this analysis also provides new insights into modified spontaneous emission in the cavity QED framework. Our approach greatly improves the quality of single photons emitted from quantum emitters even at room temperature. We expect a similar scheme can significantly improve the indistinguishability of single photons at low temperature, which is important to increase the fidelity of heralded photon-mediated two-qubit gates.

\begin{acknowledgments}
H.C. was supported in part by a Samsung Scholarship and the Air Force Office of Scientific Research (AFOSR) MURI on Optimal Quantum Measurements and State Verification. D.Z. was supported by the National Science Scholarship from A*STAR, Singapore. Y.Y. was supported in part by Skoltech as part of the Skoltech-MIT Next Generation Program. D.E. acknowledges partial support from the DARPA SEQUOIA program and the AFOSR PECASE program, supervised by Dr. Gernot Pomrenke. We thank Kevin A. Fischer for helpful discussion.
\end{acknowledgments}

\bibliographystyle{apsrev4-1}

\end{document}


\title{Supplemental Material: Indistinguishable Single Photon Sources with Dissipative Emitter Coupled to Cascaded Cavities}
\date{\today}

\maketitle

\renewcommand{\theequation}{S\arabic{equation}}
\renewcommand{\thefigure}{S\arabic{figure}}
\renewcommand{\bibnumfmt}[1]{[S#1]}
\renewcommand{\citenumfont}[1]{S#1}
\newcommand{\Lagr}{\mathcal{L}}

\section{The master equation and optical Bloch equation of the system}
Here, we write the master equation and the optical Bloch equation of the system where the quantum emitter is coupled to the cascaded cavity. The Hamiltonian of the system after the rotating wave approximation~\cite{Knight.Rotating} is,
\begin{align}
H = & \hbar \omega_{e}\hat{e}^\dagger \hat{e}+ \hbar \omega_{a} \hat{a}^\dagger \hat{a} + \hbar \omega_{b} \hat{b}^\dagger \hat{b} \nonumber
\\ & + \hbar g_1 (\hat{e}\hat{a}^\dagger+\hat{e}^\dagger \hat{a}) + \hbar g_2 (\hat{a}\hat{b}^\dagger+\hat{a}^\dagger \hat{b}) 
\end{align}
where $\hat{e}$ is the fermionic annihilation operator for the quantum emitter, and $\hat{a}, \hat{b}$ are the bosonic annihilation operators for the first and second cavity, $\omega_{e, a, b}$ are the resonance frequency of the emitter and two cavities respectively, and $g_{1,2}$ are the emitter-cavity and cavity-cavity coupling strengths. We assume the optical pumping faster than the emitter decay, which is not possible for the plasmonic approach (see the main text for the details). Thus, initial state is assumed to be $\ket{e,0,0}$ where the first index represent the emitter state (excited state), and the successive do the photon number in the cavities. System dynamics is limited in the state space $\left\{\ket{e,0,0}, \ket{g,1,0}, \ket{g,0,1}, \ket{g,0,0} \right\}$. We write the Hamiltonian in the subspace, $\left\{\ket{e,0,0}, \ket{g,1,0}, \ket{g,0,1}\right\}$,
\begin{align}
H = \hbar
\begin{bmatrix} 
0 & g_1 & 0 \\ g_1 & 0 & g_2  \\  0 & g_2 & 0
\end{bmatrix}
,
\end{align}
where the rotating frame is used with frequency $\omega_e=\omega_a=\omega_b$. States in the Hilbert space decay into $\ket{g,0,0}$ by Lindblad superoperators, and this Lindbladian can be expressed in the same basis by,
\begin{align}
\Lagr_{e}  & = -\gamma 
\begin{bmatrix} 
\rho_{ee} & \rho_{ea}/2 & \rho_{eb}/2 \\ \rho_{ae}/2 & 0 & 0  \\  \rho_{be}/2 &0&0 
\end{bmatrix}
,
\\ \Lagr_{a} & = -\kappa_1
\begin{bmatrix} 
0&\rho_{ea}/2&0\\\rho_{ae}/2&\rho_{aa}&\rho_{ab}/2\\0&\rho_{ba}/2&0 
\end{bmatrix}
,
\\ \Lagr_{b} & = -\kappa_2
\begin{bmatrix} 
0&0&\rho_{eb}/2\\0&0&\rho_{ab}/2\\\rho_{be}/2&\rho_{ba}/2&\rho_{bb}
\end{bmatrix}
,
\end{align}
where $\Lagr_{e,a,b}$ represent the decay of the emitter, first cavity ($C_1$) and the second cavity ($C_2$), respectively. In addition, we consider the Lindbladian for pure-dephasing processes.
\begin{align}
 \Lagr_{e\text{.deph}}  & = -\gamma^*
\begin{bmatrix}
0 & \rho_{ea}/2 &  \rho_{eb}/2 \\ \rho_{ae}/2 & 0 & 0 \\  \rho_{be}/2 & 0 & 0
\end{bmatrix}
.
\end{align}

The system evolves with the equation of motion,
\begin{align}
\frac{\partial \rho}{\partial t} = \Lagr[\rho] = \frac{i}{\hbar}[\rho,H] + \Lagr_{e} + \Lagr_{e\text{.deph}} + \Lagr_{a} + \Lagr_{b}. \label{master equation}
\end{align}
We rewrite Eq.~\ref{master equation} in terms of the elements of the density matrix,
\begin{align}
\frac{\partial \rho_{ee}}{\partial t} &= -\gamma\rho_{ee}+ig_1(\rho_{ea}-\rho_{ae}), \\
\frac{\partial \rho_{aa}}{\partial t} &= -\kappa_1\rho_{aa}+ig_1(\rho_{ae}-\rho_{ea})+ig_2(\rho_{ab}-\rho_{ba}), \\
\frac{\partial \rho_{bb}}{\partial t} &= -\kappa_2\rho_{bb}+ig_2(\rho_{ba}-\rho_{ab}), \\
\frac{\partial \rho_{ea}}{\partial t} &= -\frac{\gamma+\gamma^*+\kappa_1}{2}\rho_{ea}+ig_1(\rho_{ee}-\rho_{aa}) + ig_2\rho_{eb}, \label{OBE_rho_ea}\\
\frac{\partial \rho_{eb}}{\partial t} &= -\frac{\gamma+\gamma^*+\kappa_2}{2}\rho_{eb}+i(g_2\rho_{ea}-g_1\rho_{ab}), \label{OBE_rho_eb}\\
\frac{\partial \rho_{ab}}{\partial t} &= -\frac{\kappa_1+\kappa_2}{2}\rho_{ab}+ig_2(\rho_{aa}-\rho_{bb}) -ig_1\rho_{eb} \label{OBE_rho_ab},
\end{align}
which is the extended form of the optical Bloch equation.

\section{Adiabatic elimination and the rate equation}
We consider highly dissipative quantum emitter at room temperature where $\gamma^*$ is larger than any other dissipation. Thus, for $t>1/\gamma^*$, the coherence between the emitter and the photon in $C_2$ can be adiabatically eliminated ($\frac{\partial\rho_{eb}}{\partial t}=0$), and from the Eq.~\ref{OBE_rho_eb},
\begin{align}
\rho_{eb} = \frac{2i(g_2\rho_{ea}-g_1\rho_{ab})}{\gamma+\gamma^*+\kappa_2}.
\end{align}
Putting this into Eq.~\ref{OBE_rho_ea} and Eq.~\ref{OBE_rho_ab} (equations for the other two coherence terms) gives,
\begin{align}
\frac{\partial \rho_{ea}}{\partial t} &= -\frac{\gamma+\gamma^*+\kappa_1}{2}\rho_{ea}+ig_1(\rho_{ee}-\rho_{aa}) - \frac{2g_2(g_2\rho_{ea}-g_1\rho_{ab})}{\gamma+\gamma^*+\kappa_2}, \label{rho_ea_before_AE}\\
\frac{\partial \rho_{ab}}{\partial t} &= -\frac{\kappa_1+\kappa_2}{2}\rho_{ab}+ig_2(\rho_{aa}-\rho_{bb}) + \frac{2g_1(g_2\rho_{ea}-g_1\rho_{ab})}{\gamma+\gamma^*+\kappa_2}. \label{rho_ab_before_AE}
\end{align}
Introducing the population transfer rate, $R_1 = 4g_1^2/(\gamma+\gamma^*+\kappa_1)$, Eq.~\ref{rho_ab_before_AE} gives,
\begin{align}
\frac{\partial \rho_{ab}}{\partial t} \approx -\frac{\kappa_1+\kappa_2+R_1}{2} \rho_{ab} +ig_2(\rho_{aa}-\rho_{bb}),
\end{align}
where $\frac{2g_1g_2\rho_{ea}}{\gamma+\gamma^*+\kappa_2}$ term is neglected because $g_2\rho_{ea}\ll g_1\rho_{ab}$ for $t>1/R_1$, resulting in self-consistent solution, as we will see below. Because we are interested in the photon in $C_2$, which can be built up for $t>1/R_1$, $\frac{\partial \rho_{ab}}{\partial t}$ is adiabatically eliminated,
\begin{align}
\rho_{ab} = \frac{2ig_2(\rho_{aa}-\rho_{bb})}{\kappa_1+\kappa_2+R_1}.
\end{align}

Coming back to the Eq.~\ref{rho_ea_before_AE}, 
\begin{align}
\frac{\partial \rho_{ea}}{\partial t} =& -\frac{\gamma+\gamma^*+\kappa_1+4g_2^2/(\gamma+\gamma^*+\kappa_2)}{2}\rho_{ea}\\ \nonumber
& +ig_1(\rho_{ee}-\rho_{aa}) + \frac{4ig_1g_2^2(\rho_{aa}-\rho_{bb})}{(\gamma+\gamma^*+\kappa_2)(\kappa_1+\kappa_2+R_1)}\\
\approx& -\frac{\gamma+\gamma^*+\kappa_1}{2}\rho_{ea}+ig_1(\rho_{ee}-\rho_{aa}),
\end{align}
where the equation is approximated by ignoring $\Theta(1/\gamma^*)$ terms. Again by the adiabatic elimination for $t>1/\gamma^*$,
\begin{align}
\rho_{ea} = \frac{2ig_1(\rho_{ee}-\rho_{aa})}{\gamma+\gamma^*+\kappa_1}.
\end{align}
In essence, all the coherence terms are adiabatically eliminated, and coherences follow the population. Thus, the system dynamics can be described by rate equations.
\begin{align}
\dot{P_e} = \frac{\partial \rho_{ee}}{\partial t} =& -\gamma\rho_{ee}-\frac{4g_1^2}{\gamma+\gamma^*+\kappa_1}(\rho_{ee}-\rho_{aa}),\\
\dot{P_a} = \frac{\partial \rho_{aa}}{\partial t} =& -\kappa_1\rho_{aa}-\frac{4g_1^2}{\gamma+\gamma^*+\kappa_1}(\rho_{aa}-\rho_{ee})\nonumber \\ 
&-\frac{4g_2^2}{\kappa_1+\kappa_2+R_1}(\rho_{aa}-\rho_{bb}),\\
\dot{P_b} = \frac{\partial \rho_{bb}}{\partial t} =& -\kappa_2\rho_{bb}-\frac{4g_2^2}{\kappa_1+\kappa_2+R_1}(\rho_{bb}-\rho_{aa}),
\end{align}
where $P_e, P_a, P_b$ are the population of the emitter, $C_1$ and $C_2$.

Introducing the population transfer rate between two cavities, $R_2 = \frac{4g_2^2}{\kappa_1+\kappa_2+R_1}$ and converting into the matrix form,
\begin{align}
\begin{bmatrix} 
\dot{P_{e}}\\ \dot{P_{a}}\\ \dot{P_{b}}
\end{bmatrix}
&=
\begin{bmatrix} 
-\gamma-R_1 & R_1 & 0\\ R_1 & -\kappa_1-R_1-R_2 & R_2 \\0&R_2&-\kappa_2-R_2
\end{bmatrix}
\begin{bmatrix} 
P_{e}\\P_{a}\\P_{b}
\end{bmatrix}
\nonumber
\\ &= \boldsymbol{R}
\begin{bmatrix} 
P_{e}\\P_{a}\\P_{b}
\end{bmatrix}
. \label{Rate Equation}
\end{align}

\section{Comparison of simulation result of master equation and rate equation}
Here, we compare the population dynamics of the rate equation, Eq.~\ref{Rate Equation}, derived in the previous section using successive adiabatic elimination, with that of the master equation, Eq.~\ref{master equation}. We specifically used the parameters of Reg. 1 and Reg. 2 of the system in the main text. Figure S1 confirms the validity of our description with a set of rate equations (see the caption for the detailed parameters).
\begin{figure}[ht]
\includegraphics[width = \columnwidth, trim=10 0 0 10,clip]{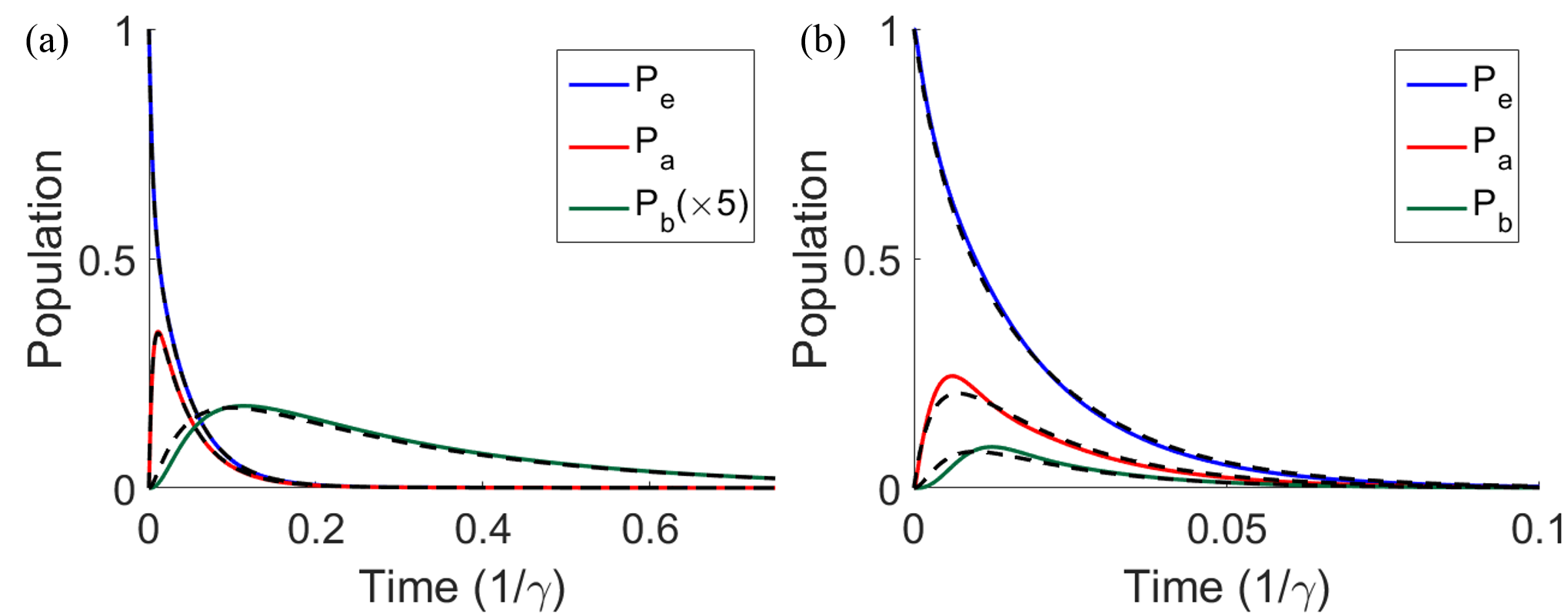}
\caption{Master equation simulation (solid) vs. rate equation simulation (dashed). (a) In Reg. 1, $g_1 = 500$, $\kappa_1 = 50$, $g_2 = 10$ and $\kappa_2 = 1$. The population of $C_2$ ($P_b$) is magnified by 5 times for clarity. (b) In Reg. 2, $g_1 = 500$, $\kappa_1 = 50$, $g_2 = 150$ and $\kappa_2 = 300$.}
\end{figure}

\section{Derivation of the efficiency}
Here, we derive the photon collection efficiency of the system through the $C_2$. Characteristic equation of the matrix $\boldsymbol{R}$ is,
\begin{align}
&(s+\gamma+R_1)\left[(s+\kappa_1+R_1+R_2)(s+\kappa_2+R_2)-R_2^2\right] \nonumber \\
&-R_1^2(s+\kappa_2+R_2)=0. \label{char_eq_efficiency} 
\end{align}
Three solutions of Eq.~\ref{char_eq_efficiency} are $s=-s_1,-s_2,-s_3$. Then, $P_b=Ae^{-s_1t}+Be^{-s_2t}+Ce^{-s_3t}$. $A$, $B$ and $C$ can be expressed in terms of $s_1, s_2, s_3$ using initial conditions, $P_b(0)=0, \dot{P_b}(0)=0, \ddot{P_b}(0)=R_1R_2$. The efficiency is,
\begin{align}
\eta = \kappa_2\int^\infty_0 P_b(t) dt = \kappa_2 \left(\frac{A}{s_1}+\frac{B}{s_2}+\frac{C}{s_3}\right) = \kappa_2\frac{R_1R_2}{s_1s_2s_3}.
\end{align}
The denominator is the coefficient of the constant term of the Eq.~\ref{char_eq_efficiency} with the sign reversed. Therefore,
\begin{align}
\eta =  \frac{\kappa_2R_2}{\kappa_1(\kappa_2+R_2)+\kappa_2R_2}.
\end{align}
Note that this method can be extended to arbitrarily large systems with the dynamics described by rate equations.

\section{Effective Emitter Method}
As we will show in the next section, deriving indistinguishability requires the decay rate of the photon in $C_2$. However, direct solution of Eq.~\ref{char_eq_efficiency} has a complex form with cube root resulting in vague physical interpretation. Here, we formally describe the system by constructing effective (composite) emitter out of the emitter and the photonic mode of $C_1$, and derive the decay constant of the $C_2$-like eigenmode.

The system matrix of the rate equation, Eq.~\ref{Rate Equation}, is transformed by the matrix T,
\begin{widetext}
\begin{align}
T &= 
\begin{bmatrix}
1 & 1 & 0 \\
\frac{2R_1}{(\kappa_1+R_2)-\sqrt{(\kappa_1+R_2)^2+4R_1^2}} & \frac{2R_1}{(\kappa_1+R_2)+\sqrt{(\kappa_1+R_2)^2+4R_1^2}} & 0 \\
0 & 0 & 1
\end{bmatrix}
\approx
\begin{bmatrix}
1 & 1 & 0 \\
-1 & 1 & 0 \\
0 & 0 & 1
\end{bmatrix}
\text{(for } 2R_1\gg\kappa_1+R_2\text{)}
,
\end{align}
through the transformation $[P_e, P_a, P_b]^T\rightarrow T^{-1}[P_e, P_a, P_b]^T = [P_d, P_s, P_b]^T$ and $\boldsymbol{R}\rightarrow T^{-1}\boldsymbol{R}T$. As a result,
\begin{align}\label{Eff_Eq}
\begin{bmatrix} 
\dot{P_{d}}\\ \dot{P_{s}}\\ \dot{P_{b}}
\end{bmatrix}
=
\begin{bmatrix} 
-\frac{(\kappa_1+R_2)+2R_1+\sqrt{(\kappa_1+R_2)^2+4R_1^2}}{2} & 0 & -\frac{R_1R_2}{\sqrt{(\kappa_1+R_2)^2+4R_1^2}}\\
0 & -\frac{(\kappa_1+R_2)+2R_1-\sqrt{(\kappa_1+R_2)^2+4R_1^2}}{2}&\frac{R_1R_2}{\sqrt{(\kappa_1+R_2)^2+4R_1^2}}\\
-\frac{2R_1R_2}{\sqrt{(\kappa_1+R_2)^2+4R_1^2}-(\kappa_1+R_2)}&\frac{2R_1R_2}{\sqrt{(\kappa_1+R_2)^2+4R_1^2}+(\kappa_1+R_2)}&-\kappa_2-R_2
\end{bmatrix}
\begin{bmatrix} 
P_{d}\\P_{s}\\P_{b}
\end{bmatrix}
,
\end{align}
\end{widetext}
where $\gamma$ ($\ll R_1$) is neglected. When $2R_1 \gg \kappa_1+R_2$, Eq.~\ref{Eff_Eq} can be reduced to,
\begin{align}
\begin{bmatrix} 
\dot{P_{d}}\\ \dot{P_{s}}\\ \dot{P_{b}}
\end{bmatrix}
=
\begin{bmatrix} 
-(2R_1+\frac{\kappa_1+R_2}{2}) & 0 & -\frac{R_2}{2}\\
0 & -\frac{\kappa_1+R_2}{2}&\frac{R_2}{2}\\
-R_2&R_2&-(\kappa_2+R_2)
\end{bmatrix}
\begin{bmatrix} 
P_{d}\\P_{s}\\P_{b}
\end{bmatrix}
.
\end{align}

\begin{figure}[ht]
\includegraphics[width = \columnwidth, trim=5 5 5 5,clip]{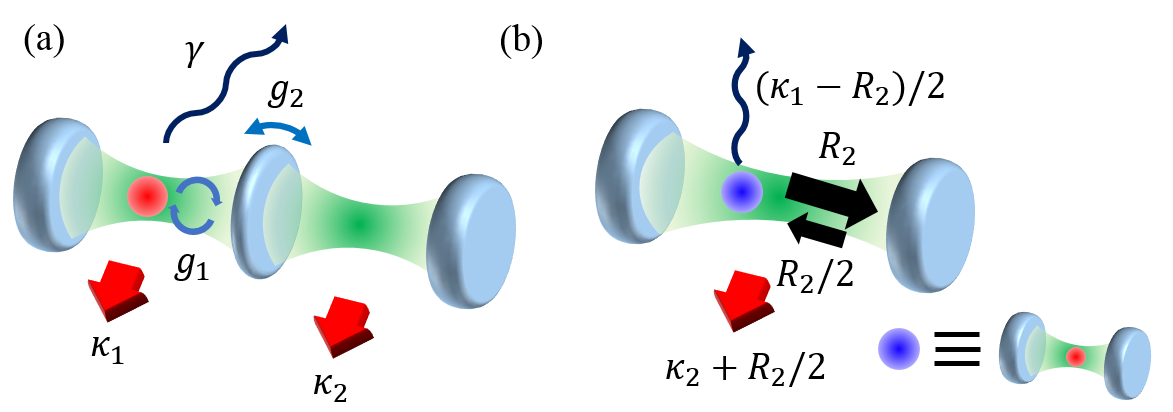}
\caption{(a) Cascaded cavity coupled to the emitter. (b) Effective--emitter picture. Blue sphere represents effective emitter $P_s$. This composite emitter is coupled to the $C_2$ with asymmetric population transfer rate $R_2$ and $R_2/2$.}\label{effective emitter picture}
\end{figure}

Here, we made an approximation that $P_d$ does not affect the long-timescale behavior of $P_b$, i.e. the decay rate of it. This approximation is based on the large difference between diagonal terms : $2R_1+(\kappa_1+R_2)/2\gg \kappa_2+R_2$. The fractional error of this approximation on the decay constant is $\sim O(R_2^2/(2R_1+\kappa_1/2-\kappa_2-R_2/2)(\kappa_2+R_2))$, which on the order of $\sim10^{-2}$ in our example (Regime 1). In other words, we can set $P_d(t)=0$ for $t>1/2R_1$. As a result, we can describe the system with two variable $P_s$ and $P_b$ coupled by $2\times2$ matrix.
\begin{align}
\begin{bmatrix} 
\dot{P_{s}}\\ \dot{P_{b}}
\end{bmatrix}
=
\begin{bmatrix} 
-\frac{\kappa_1+R_2}{2}&\frac{R_2}{2}\\
R_2&-(\kappa_2+R_2)
\end{bmatrix}
\begin{bmatrix} 
P_{s}\\P_{b}
\end{bmatrix}
. \label{effective emitter rate eq}
\end{align}
Physically, Eq.~\ref{effective emitter rate eq} describe the composite emitter ($P_s$), which is the sum of population of $P_e$ and $P_a$, is coupled with $C_2$. Importantly, the coupling is asymmetric, and $P_s$ pumps $P_b$ with rate $R_2$ while the reverse happens with the rate $R_2/2$. Figure \ref{effective emitter picture} illustrates effective emitter description based on Eq.~\ref{effective emitter rate eq}. From Eq.~\ref{effective emitter rate eq}, we calculated the decay rate of photonic excitation in $C_2$,
\begin{align}\label{effective_emitter_Pb}
P_b \sim \text{exp}\left(-\frac{\kappa_1(\kappa_2+R_2)+\kappa_2R_2}{\kappa_1+2\kappa_2+3R_2}t\right).
\end{align}

\section{Indistinguishability of Regime 1}
We use the non-equilibrium Green's function formalism and define retarded green's function~\cite{Haug.NEGF},
\begin{widetext}
\begin{align}
\hat{G}^R(\tau) = \Theta(\tau)
\begin{bmatrix}
\left<\hat{e}^\dagger(t+\tau)\hat{e}(t)+\hat{e}(t)\hat{e}^\dagger(t+\tau)\right>		
&\left<\hat{e}^\dagger(t+\tau)\hat{a}(t)+\hat{a}(t)\hat{e}^\dagger(t+\tau)\right>			
&\left<\hat{e}^\dagger(t+\tau)\hat{b}(t)+\hat{b}(t)\hat{e}^\dagger(t+\tau)\right>
\\ \left<\hat{a}^\dagger(t+\tau)\hat{e}(t)+\hat{e}(t)\hat{a}^\dagger(t+\tau)\right>
&\left<\hat{a}^\dagger(t+\tau)\hat{a}(t)+\hat{a}(t)\hat{a}^\dagger(t+\tau)\right>
&\left<\hat{a}^\dagger(t+\tau)\hat{b}(t)+\hat{b}(t)\hat{a}^\dagger(t+\tau)\right>
\\\left<\hat{b}^\dagger(t+\tau)\hat{e}(t)+\hat{e}(t)\hat{b}^\dagger(t+\tau)\right>
&\left<\hat{b}^\dagger(t+\tau)\hat{a}(t)+\hat{a}(t)\hat{b}^\dagger(t+\tau)\right>
&\left<\hat{b}^\dagger(t+\tau)\hat{b}(t)+\hat{b}(t)\hat{b}^\dagger(t+\tau)\right>
\end{bmatrix}
.
\end{align}
\end{widetext}
With the Markovian self-energies that we assumed in the first section, the Dyson's equation gives,
\begin{align}
\hat{G}^R(\omega) = 
\begin{bmatrix}
\omega+i\gamma/2+i\gamma^*/2 	 &g_1 					 & 0 \\
g_1							 &\omega+i\kappa_1/2		 & g_2 \\
0 							 &g_2 					 & \omega+i\kappa_2/2 
\end{bmatrix}
^{-1}.
\end{align}
One can readily express the two-point correlation function with retarded green's function under Markovianity~\cite{Thomas.Funnel},
\begin{align}
\left< \hat{b}^\dagger(t+\tau)\hat{b}(t) \right> = \bra{g,0,1}\hat{G}^R(\tau)\hat{\rho}(t)\ket{g,0,1}
\\ = G_{bb}^R(\tau)\rho_{bb}(t)+G_{ba}^R(\tau)\rho_{ab}(t)+G_{be}^R(\tau)\rho_{eb}(t).
\end{align}
Note that the physics of this step is the same with the quantum regression theorem, which also has been used to study the indistinguishability of the dissipative cQED system~\cite{Simon.Plasmon}.

In Reg. 1, $\gamma+\gamma^*-\kappa_1\gg g_1$ and $R_1+\kappa_1-\kappa_2\gg g_2$. Therefore, by using binomial expansion to the solution of the characteristic equation,
\begin{align}
G_{be}(\tau) &\sim e^{-\frac{\gamma+\gamma^*-R_1}{2}\tau} \\
G_{ba}(\tau) &\sim e^{-\frac{\kappa_1+R_1-R_2}{2}\tau} \\
G_{bb}(\tau) &\sim e^{-\frac{\kappa_2+R_2}{2}\tau}.
\end{align}
Thus, for $\tau\sim1/(\kappa_2+R_2)$, $\left< \hat{b}^\dagger(t+\tau)\hat{b}(t) \right> \sim G_{bb}^R(\tau)\rho_{bb}(t)$.

On the other hand, indistinguishability ($I$) is defined by Hong-Ou-Mandel interference visibility~\cite{Imamoglu.Ind},
\begin{align}
I = 1 - 2p_c = \frac{\int^\infty_0dt\int^\infty_0d\tau\abs{\left< \hat{b}^\dagger(t+\tau)\hat{b}(t) \right>}^2}{\int^\infty_0dt\int^\infty_0d\tau\left< \hat{b}^\dagger(t)\hat{b}(t) \right>\left< \hat{b}^\dagger(t+\tau)\hat{b}(t+\tau) \right>},
\end{align}
where $p_c$ is the coincidence probability in Hong-Ou-Mandel experiment. We can hence express $I$ as,
\begin{align}
I \approx \frac{\int^\infty_0dtP_b^2(t)\int^\infty_0d\tau G_{bb}^2(\tau)}{\frac{1}{2}\abs{\int^\infty_0dtP_b(t)}^2}.
\end{align}
By plugging Eq.~\ref{effective_emitter_Pb} in,
\begin{align}
I \approx \frac{\kappa_1/2+(\kappa_2||R_2)/2}{\kappa_1/2+\kappa_2+3R_2/2}.
\end{align}

\bibliographystyle{apsrev4-1}